\documentclass[10pt]{article}
\usepackage{setspace}
\usepackage{amssymb}
\usepackage{amsmath}
\usepackage{amsfonts}
\usepackage{bm}
\usepackage{graphicx}

\def\be{\begin{equation}}

\def\ee{\end{equation}}

\newcommand\of[1]{\left( #1 \right)}

\def\bea{\begin{eqnarray}}

\def\eea{\end{eqnarray}}

\newcommand\refeq[1]{(\ref{#1})}
\begin{document}

\singlespace

\vspace*{.3in}

\begin{center}

{\Large\bf Approximate Born-Infeld Effects on the Relativistic Hydrogen Spectrum}

{\large J.\ Franklin} \\
{\it  Reed College, Portland, OR 97202 \\
{\tt jfrankli@reed.edu}}

{\large T.\ Garon} \\
{\it  Reed College, Portland, OR 97202 \\
}

\end{center}

\begin{abstract}
The Born-Infeld form of the hydrogen atom has a spectrum that can be used to determine the physical viability of the theory, and place an experimentally relevant bound on the single parameter found in it.  We compute this spectrum using the relativistic Dirac equation, and a form of the Born-Infeld potential that approximates the self-field corrections of the electron.  Using these together, we can establish that if the Born-Infeld nonlinear electrodynamics is to be physically relevant, it must contain a fundamental constant that is well-below the original value proposed by Born.  This work extends the original Schr\"odinger spectrum from \cite{PhysRevLett.96.030402} for the self-field correction, and shows that using the Dirac equation introduces minor corrections -- but also gives access to a range for the fundamental constant that is below that attainable from non-relativistic considerations.

\end{abstract}

\section{Introduction}

The motivations for studying Born-Infeld electrodynamics~\cite{M.Born03291934,M.Born1933} are varied -- while the original motivation (finite point-source self-energy) may or may not be compelling in modern times, the utility of the theory as, at the very least, a proving ground for potential nonlinear modification is not to be undervalued \cite{ANN,PhysRevE.77.046601,PhysRevLett.99.230401}.  Of particular interest here is the role of particles, which both source and respond to nonlinear fields \cite{PLA,JSP.Kiessling.1}.  In addition to providing a vehicle for thinking carefully about nonlinearity (in preparation for other nonlinear studies, like GR, say \cite{BIG}), the particulars of Born-Infeld have current proponents from string theory (by now, this is established enough to warrant a chapter in \cite{Zwiebach}), general relativity, atomic physics \cite{PhysRevLett.27.958, PhysRevA.7.903} and spectroscopy \cite{Optics.Spectroscopy}.   Our work serves both as an attempt to collect and compare known spectra, as well as present the results of the Dirac spectrum, all in a unified language.  The ultimate goal of any such spectral study is to place a bound on the parameter governing the strength of the Born-Infeld potential.  This single number, the only input in the theory, can be used to evaluate the physical relevance of the Born-Infeld approach, in addition to setting the range of applicability of the theory.

The Born-Infeld field equation for the electrostatic potential is:
\begin{equation}
\nabla \cdot \left[ \frac{\nabla V}{\sqrt{1 - \frac{\of{\nabla V}^2 \, \of{\tilde a \, a}^2}{e/(4 \, \pi \, \epsilon_0)}}}\right] = - \frac{\rho}{\epsilon_0}
\end{equation}
where $a$ is the Bohr radius of hydrogen, $\tilde a$ is a parameter that sets the value of the electrostatic energy density at $r = 0$ and $e$ is the charge of the electron.  Born proposed a value of $\tilde a_B = \frac{1}{6} \, B(1/4, 1/4) \, \alpha^2$ (where $\alpha = 7.2973525376\times10^{-3}$, the fine structure constant, and $B$ is Euler's $\beta$ function) that comes from equating the (now finite) field energy with the rest energy of the electron (the actual number is usually given with dimensions of length, $a_B = \tilde a_B \, a$).  If we set the density on the right to $\rho = e \, \delta^3({\bf r})$, corresponding to a point particle, we recover the usual expression for the potential energy of a charge $-e$ moving in the field of a central charge $e$:
\begin{equation}\label{BIV}
V^1_{BI}(r) = -\frac{e^2}{4 \, \pi \, \epsilon_0} \, \frac{1}{\tilde a\, a} \, \int_{r/(\tilde a \, a)}^\infty \frac{dx}{\sqrt{1 + x^4}}.
\end{equation}

Previous work on spectra has focused on this ``test-particle" approximation (as in \cite{TPARTH,Optics.Spectroscopy}), where only the single-particle Born-Infeld point potential is used.  This approach is justified in the limit that the proton charge is much larger than the electron's, so that the electron's contribution to the total Born-Infeld field is negligible.  Of course, for hydrogen, this approximation is invalid, but the test particle potential is still useful.  The Born-Infeld field equations are nonlinear, and so the dipole field (of a proton and an electron) is not the sum of the individual point fields.  It is this dipole field, associated with the density $\rho = e\, \delta^3({\bf r}-{\bf r}_+) - e \, \delta^3({\bf r} - {\bf r}_-)$, that one would ultimately like to use as a potential. In three dimensions, the dipole field has no closed form, and so the test particle potential is relevant to spectral studies as the only known exact particle solution.

An approximation to the dipole potential was proposed \cite{JSP.Kiessling.2} that includes some of the self-field effects of the electron.  In \cite{PhysRevLett.96.030402}, the non-relativistic Schr\"odinger spectrum was calculated with this potential, and it was shown that in this approximation, the parameter $\tilde a$ in the Born-Infeld equation must be less than Born's original value.  In this paper, we extend the spectrum to include relativistic effects by computing the spectrum of hydrogen, with this modified potential, using the Dirac equation.  By including relativistic effects, we can probe values of the parameter $\tilde a$ below the original $\tilde a_B$ given by Born -- when we compare the spectra obtained by the Schr\"odinger and Dirac equations, the differences between the test-particle potential and the self-field potential appear at roughly an order of magnitude larger than the relativistic corrections for $\tilde a =\tilde a_B$.  Using the Dirac equation allows us to distinguish between spectral differences between the two potentials, and those arising from relativistic effects for values of $\tilde a$ well below $\tilde a_B$.  In addition to corrections to the energies themselves, the Dirac spectrum will provide angular degeneracy information that can be used to further constrain the allowed values of $\tilde a$.

We will show that as far as the test-particle form is concerned (relevant if we had a system in which the ``electron" had charge much less than the ``proton"), deviation of spectra (in all cases) from accepted values is minimal for ``reasonable" values of the parameter found in the Born-Infeld potential (i.e.\ near the one proposed by Born).  If one takes into account the electron self-field, then parametric values are constrained by angular degeneracy (this was shown for the non-relativistic case in~\cite{PhysRevLett.96.030402}).  After setting up the physical framework (the potentials), we discuss the numerical method used to find all spectra in this paper, the simplest method possible.  Then we set the two parameters of the method, and indicate its limitations, by matching the Dirac hydrogen spectrum (using the Coulomb potential).  
From there, we are ready to calculate all flavors of spectra, and finally, show that in the self-field case, the angular portion of the Dirac spectra would be measurably different if BI holds, even for values of the BI parameter down to one-tenth of the theoretical limit.  It is important to note that this self-field potential is still approximate -- in order to treat the full hydrogen problem, one would need a three-dimensional dipole solution to the Born-Infeld field equations, absent a fully quantum form for the field.

\section{Setup}

Here, we briefly review the self-field potential found in \cite{JSP.Kiessling.2}, and set the form and dimensionless variables appropriate for both Schr\"odinger and Dirac investigations.

The original Born-Infeld potential energy for a particle of charge $e$ interacting with a test particle of charge $-e$ was given in~\refeq{BIV}.
Kiessling has argued~\cite{JSP.Kiessling.1} that when considering a two-body system like hydrogen, we must include the self-field of the electron, in addition to the central charge, and gives an augmented potential, approximating this correction, of the form:
\begin{equation}\label{KV}
\begin{aligned}
V^2_{BI}(r) &= -\frac{e^2}{4 \, \pi \, \epsilon_0} \, \frac{1}{\tilde a \, a} \, \biggl[ \frac{r}{\tilde a \, a}\,  \int_0^{1/(2 \, \sqrt{2})} \frac{ 2 \, x \, \sqrt{1 + x^2} - 2 \, x^2 - 1}{\sqrt{1 + 4 \, x^2 - 4 \, x \, \sqrt{1 + x^2}} \, \sqrt{1 + x^2} \, \sqrt{1 + \of{\frac{r}{\tilde a \, a}}^4 \, x^4} } \, dx \\
&+ \frac{1}{4} \, B(1/4,1/4) \biggr]. 
\end{aligned}
\end{equation}
We have used the identifiers $1$ and $2$ in~\refeq{BIV} and~\refeq{KV} to distinguish between these two potentials, and remind us of the number of charges being approximated.

For Schr\"odinger's equation, we start from the separated radial ODE, with $u(r) = r \, R(r)$ a function of $r$ only.  If we set $r = \rho \, a$ for Bohr radius $a$ and dimensionless $\rho$, then the radial equation can be written:
\begin{equation}\label{Sch}
-u'' + \left[ 2 \, V(\rho) + \frac{\ell \, \of{\ell+1}}{\rho^2}\right] \, u= \frac{2}{\alpha^2} \, \frac{E}{m \, c^2} \, u
\end{equation}
with $u'' \equiv \frac{d^2 u}{d\rho^2}$.  The energy on the right,  for the Coulomb potential, would be $\frac{2}{\alpha^2} \, \frac{E}{m \, c^2} = -\frac{1}{n^2}$.   In these variables, the (test-particle) Born-Infeld potential for use in Schr\"odinger's equation is:
\begin{equation}
V_{BI}^1(\rho) = -\frac{1}{\tilde a} \, \int_{\rho/\tilde a}^\infty \frac{dx}{\sqrt{1 + x^4}}, 
\end{equation}
with similar modification for the potential~\refeq{KV}.

Using the same dimensionless $\rho$, the Dirac equation can be brought to the form:
\begin{equation}
\begin{aligned}
u' + \frac{\kappa}{\rho} \, u + \of{\alpha \, V(\rho)- \frac{1}{\alpha}} \, v &= \frac{E}{\alpha \, m \, c^2} \, v \\
-v' + \frac{\kappa}{\rho} \, v + \of{\alpha \, V(\rho) + \frac{1}{\alpha}} \, u &= \frac{E}{\alpha \, m \, c^2} \, u
\end{aligned}
\end{equation}
where $u$ and $v$ make up the radial portion of the full spinor.  We define $\kappa \equiv j + \frac{1}{2}$ for $j$ the total angular momentum.  

\section{Method}

We use the same numerical method to find the spectra in all cases -- finite differences on a regular grid with $\rho_j = j \, \Delta \rho$ (for some choice of step $\Delta \rho$). 
A centered difference replaces the second derivative $u''$ in the Schr\"odinger equation, and a similar centered difference is applied to the first derivative(s) appearing in the Dirac equation.

In particular, the second derivative $u''$ appearing in Schr\"odinger's equation can be written approximately as
\begin{equation}
u''(\rho_j) \approx \frac{u_{j+1} - 2 \, u_j + u_{j-1}}{\Delta \rho^2} + O(\Delta \rho^2)
\end{equation}
so that the second derivative of $u$ at the grid point $\rho_j$ consists of a combination of the values of the function $u$ at $\rho_{j+1}$, $\rho_j$ and $\rho_{j-1}$.  While it is clear how to embed this information in a matrix for most grid-points, the ``boundary points" (like $\rho_1$) are more subtle.  Our grid starts at $\rho_1 \equiv \Delta \rho$, suppose it ends at $\rho_N \equiv N \, \Delta \rho$ for some integer $N$.  We will need to refer to the values of the function $u$ at zero and $(N+1) \, \Delta \rho$ to construct the approximation to the second derivative of $u$ at $\rho_1$ and $\rho_N$.  At the left-hand boundary, the situation is straightforward -- the $u$ in~\refeq{Sch} is $r \, R(r)$ for $R(r)$ the radial portion of the wavefunction.  For hydrogenic $R_{n\ell}(r)$, we know that $R_{n\ell}(0)$ is zero or finite, so $u(0) = 0$ automatically.  That means we can ignore the reference to $u_0$ appearing in the finite difference, since it is zero.

The other end is more difficult -- we expect all $u$ to go to zero as $\rho$ gets ``large" (compared to unity), so we choose an artificial infinity, called $\rho_\infty$, and demand that the radial portion of the wavefunction vanish at this necessarily finite value.  This approximation is not so bad for low-lying energy states, where we expect decaying exponentials in $\rho$ to dominate the behavior of $u$ for large values of $\rho$.  Still, we must be aware that higher energy states will have greater error associated with our choice.  For now, we include our choice of $\rho_\infty$ and $N$, the number of grid points, as inputs in our method, and then define $\Delta \rho \equiv \frac{\rho_\infty}{N}$ as the step size.

These approximations render the ODEs as algebraic eigenvalue problems of the form:
\begin{equation}\label{algebraic}
\mathbb{D} \, {\bm u} = \tilde E \, {\bm u},
\end{equation}
for the vector ${\bm u}$ with entries $u_j \approx u(\rho_j)$.  Our goal, then, is to solve for the energies $\tilde E$ that are just the eigenvalues of the discretized differential operator $\mathbb{D}$ (a sparse matrix).  Since we are interested in the low-end of the spectrum, we only want a few of the smallest eigenvalues, and we use the Lanczos iterative method to find just the first few eigenvalues (quickly) \footnote{In practice, we could use almost any language or numerical package to implement this method.  For simplicity, and accessibility, we used the built-in Lanczos routines found in \texttt{Mathematica}.}.
Note that an identical procedure can be applied to the first order Dirac pair, although we have to be careful to include both the $u$ and $v$ contributions correctly -- this means that ${\bm u}$ appearing in~\refeq{algebraic} is now a vector that includes all discrete values of $u$ and $v$ (a vector, then, of length $2 \, N$), and $\mathbb{D} \in {\mathbb R}^{2 N \times 2 N}$.

\subsection{Test and Parameters}

As a check of the method, and a validation of the parameters $\rho_\infty$ and $N$ used in the remainder of this paper, we reproduce the Dirac hydrogen spectrum, probing (successfully) both its principle quantum number and angular dependence  by comparing with the known spectrum (using our dimensionless variables):
\begin{equation}\label{DiracE}
\begin{aligned}
\frac{2}{\alpha}\, \frac{E}{\alpha \, m \, c^2} &=\frac{2}{\alpha^2}\,  \left\{\left[  1 +\of{ \frac{\alpha}{n - \kappa + \sqrt{\kappa^2 - \alpha^2}}}^2 \right]^{-1/2} - 1 \right\} \\
&\equiv \tilde E
\end{aligned}
\end{equation}
where, again, $\kappa = j + \frac{1}{2}$, and $n$ is the principle quantum number (for the Schr\"odinger spectrum).

Using $\rho_\infty = 100$, and $N = 20000$, we obtain a ground state energy for the Dirac equation (with the Coulomb potential: $V(\rho) = -1/\rho$) of: $\tilde E \approx -1.000013313195$ -- this compares well with the
exact value from~\refeq{DiracE} -- the energy values for the first three principle quantum numbers, and first two orbital states are shown in Table 1.  Notice, there, that the lowest energy levels ($n=1$, $\kappa=1$ and $n=2$, $\kappa=2$) in each series are the most accurate -- our direct method loses accuracy as energy increases, which is not surprising.  For the usual radial wavefunctions, higher energy implies longer spatial extent, and our designation of $\rho_\infty$ becomes relevant in the error.  In addition, the numerical spectrum is necessarily finite, so that we expect errors to accumulate near the ``top" of the spectrum, where we ultimately approximate the infinite-energy scattering states with a finite value.

\begin{center}
\begin{table}
\begin{tabular}{|l|l|l|l|}\hline
\textrm{$n$}&\textrm{$\kappa$}&\textrm{$-\tilde E$ }&\textrm{$-\tilde E$ from~\refeq{DiracE}} \\ \hline
1 & 1 & 1.000013313195 & 1.000013313195\\
2 & 1 & .2500049 & .2500042\\
3 & 1 & .1111129 &.1111126  \\
2 & 2 & .250000832055 & .250000832051\\
3 & 2 &  .11111164 & .11111160\\ \hline
\end{tabular}
\caption{Dirac spectrum calculated numerically and from~\refeq{DiracE}.  We show the numerical result up to disagreement with the theoretical values.}
\end{table}
\end{center}


\section{Schr\"odinger Spectra}

\begin{figure*}[t] 
   \centering
   \includegraphics[width=5in]{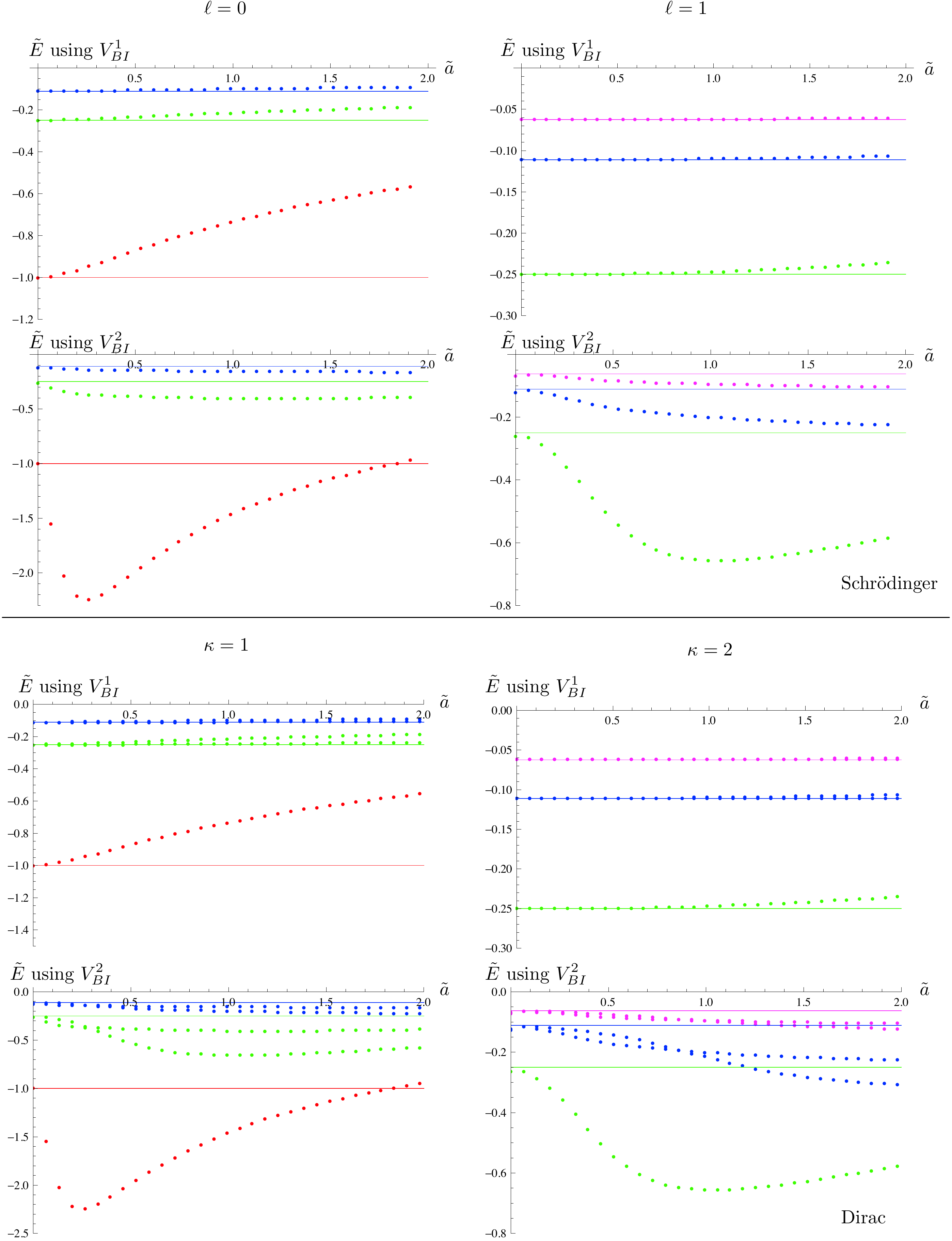} 
   \caption{Top: the energy spectra for $\ell = 0$ and $\ell = 1$ as computed using the test particle Born-Infeld potential~\refeq{BIV}, and the self-field potential~\refeq{KV}.  Solid lines indicate the Coulomb-based hydrogen spectrum.  On the bottom is the Dirac version of the same spectra, for $\kappa = 1$ and $\kappa = 2$, again computed using the test particle Born-Infeld potential, and self-field potential.   The splitting of the $\ell = 0$ and $\ell = 1$ degeneracy is evident here (for example, on the bottom right, $\kappa = 2$ case, the $n=3$ energies begin at $\tilde a = 0$ taking on their Coulomb value, but as $\tilde a$ is increased, the pair becomes split in both the test-particle and self-field potentials). }
   \label{fig:Sch}
\end{figure*}

We computed the energy spectrum as a function of $\tilde a = Q \, \tilde a_B$ for $Q = 1 \longrightarrow 30000$ in steps of
$1000$ for the ground state, and first two excited states with $\ell = 0$ and $\ell = 1$ for both the test-particle form of the potential~\refeq{BIV}, and the approximation~\refeq{KV}.  The results are shown on the top in Figure \ref{fig:Sch}.  Note that we recover the two values for $\tilde a$ that give good agreement for the hydrogen ground state energy (a hallmark of the potential~\refeq{KV}), but it is also clear that
every energy level will have two such values:  At $\tilde a = 0$, we know we recover Coulomb, and we expect that for large $\tilde a$, the test particle potential and~\refeq{KV} coincide -- but all the test-particle deviations lie above the Coulomb value, while all deviations coming from the modified potential start off below the Coulomb value.  In order for the two spectra to match for large $\tilde a$, the self-field spectrum must come back up to the Coulomb value and cross it. 

In all cases, the potential with electron field included diverges from the Coulomb spectrum more than the test particle potential.  In addition, the splitting of the angular degeneracy is significant when using~\refeq{KV} -- it is this splitting that we will probe for values of $\tilde a$ below Born's.   In Table 2, we see the numerical values associated with $\tilde a_B$ for each energy level and potential.
\begin{center}
\begin{table}
\begin{tabular}{|l|l|l|l|}\hline
\textrm{$n$}&\textrm{$\ell$}&\textrm{$-\tilde E$ for $V^2_{BI}$ }&\textrm{$-\tilde E$ for $V^1_{BI}$ } \\
\hline
1 & 0 & 1.00033 & .999994\\
2 & 0 & .25004 & .2499996\\
3 & 0 & .11112 & .11111103  \\
2 & 1 &  .250014 & .2500001 \\
3 & 1 &  .111115 & .11111117\\
4 & 1 & .06250178 & .06250003 \\ \hline
\end{tabular}
\caption{Comparison of the spectrum computed with the test-particle potential~\refeq{BIV} and the potential with electron self-field included~\refeq{KV} for Schr\"odinger's equation using $\tilde a = \tilde a_B$.}
\end{table}
\end{center}

\section{Dirac Spectra}

The general pattern for both potentials when we move to the relativistic Dirac equation is the same -- very little changes when compared to the Schr\"odinger case.  On the bottom in Figure \ref{fig:Sch}, we see the spectra for $\kappa = 1$, $2$ -- in either Born-Infeld case, the degeneracy of $\ell = 0$, $1$ is split for some value of $\tilde a$, and the splitting is more dramatic when using~\refeq{KV}.  In that case, we can see an interesting crossing in the $\kappa = 2$ case, where degeneracy is restored at some (large) value of $\tilde a$ (near $\tilde a \approx 1$).

A table of numerical values at Born's $\tilde a_B$ is shown in Table 3.  The data shown there is carried to the first digit that disagrees with the Schr\"odinger results in Table 2 for the potential~\refeq{KV}, and enough digits to distinguish between the $\ell = 0$ and $\ell = 1$ cases. 

\begin{center}
\begin{table}\label{tab:DiracKI}
\begin{tabular}{|l|l|l|l|} \hline
\textrm{$n$}&\textrm{$\kappa$}&\textrm{$-\tilde E$ for $V^2_{BI}$ }&\textrm{$-\tilde E$ for $V^1_{BI}$ } \\ \hline
1 & 1 & 1.00035 & 1.000013\\
2 & 1 & \begin{tabular}{l}.250047 \\ .250019 \end{tabular} & \begin{tabular}{l} .2500049 \\ .2500049 \end{tabular} \\
3 & 1 & \begin{tabular}{l} .111125 \\ .111117 \end{tabular} &\begin{tabular}{l}  .1111129 \\ .1111129\end{tabular} \\
2 & 2 &  .2500148 &  .25000083 \\ \hline
\end{tabular}
\caption{Comparison of the spectrum computed with the test-particle potential~\refeq{BIV} and the potential with electron self-field included~\refeq{KV} using the Dirac equation (at $\tilde a = \tilde a_B$).}
\end{table}
\end{center}

\section{Bounds on Born's Parameter}
\begin{figure}[htbp] 
   \centering
   \includegraphics[width=2.25in]{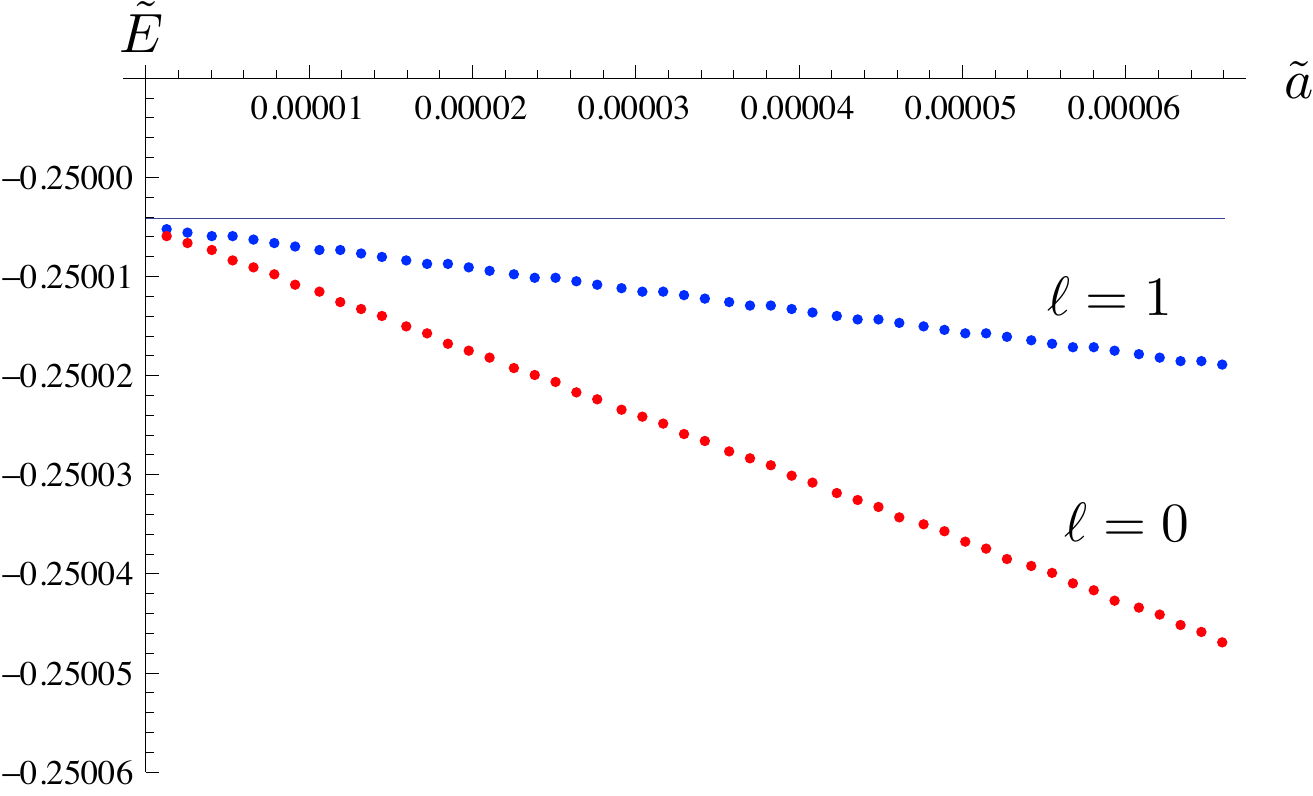} 
   \caption{The $\tilde E_2$ energy calculated with $\ell = 0$ and $\ell = 1$ for values of $\tilde a$ starting at $\tilde a_B/50$.  Here, we use the self-field potential in the Dirac equation -- the divergence of the degeneracy, for small values of parameter, indicates that even for values well below Born's, the self-field Born-Infeld potential is experimentally unacceptable.}
   \label{fig:SchLow}
\end{figure}

Using the self-field potential, we see that the spectrum of hydrogen, even at Born's modest value for $\tilde a$, is off, especially when we think of the angular degeneracy associated with $\ell = 0$ and $1$.  We probe below the Born value, to see what type of behavior ever-smaller parametric values induce.  Note that in the test particle case, all is well, so we focus on the potential~\refeq{KV}.  The result, for the Dirac equation, is shown in Figure \ref{fig:SchLow}.

\section{Conclusion}

We have demonstrated that the use of either the test-particle or modified self-field BI potential in the relativistic Dirac setting does not significantly change the Schr\"odinger results; we pick up corrections of the usual fine structure size in each case.  Using the modified potential~\refeq{KV}, we see a similar divergence in angular dependence for the hydrogen spectra for both the Schr\"odinger and Dirac equations.   But the difference between the test-particle form and the self-field form at $\tilde a_B$ is approximately an order of magnitude larger than the relativistic corrections, hence to probe values of the parameter $\tilde a$ below $\tilde a_B$, and distinguish between the two potentials, we must use a relativistic approach.  One could interpolate/extrapolate from our energies, as functions of $\tilde a$, to put concrete numerical bounds on this parameter given some experimental tolerance, or  use our results to verify or reject an externally proposed value of $\tilde a$.  Our relativistic calculation competes with the effects of spin-orbit coupling in magnitude, and it would be interesting to include some relevant Born-Infeld approximation to spin-orbit coupling in the current framework.

Our spectral results exhibit $\ell$ dependence that is numerically different than previous work 
\footnote{In~\cite{PhysRevLett.96.030402}, the energy for the $n=2$, $\ell = 0$ and $\ell = 1$ states are given as (in our current dimensionless scheme): $\tilde E_{20} = .25004$ and $\tilde E_{21} = .38202$, leading to dramatic degeneracy splitting.  We have checked our results for a variety of parameteric settings of $\rho_\infty$ and $N$, in addition to modifying the (necessarily numerical) integration of~\refeq{KV}, and in all cases (including a completely separate determination of the spectrum using shooting), we find agreement with the stated values in Table 2, values which are correctly echoed in the Dirac spectrum at $\tilde a = \tilde a_B$ as shown in Table 3.},
and imply a much milder set of spectral shifts when we use $V^2_{BI}$ -- the self-field form of the BI potential still exhibits greater splitting of angular degeneracy in both the Schr\"odinger (not shown) and Dirac settings, and this can be used to put a bound on the value of $a_B$ as needed (depending on the application, in other words).  The method we use is straightforward and generalizable to other nonlinear modifications, or physically inspired perturbations (such as spin-orbit coupling).  Future work can build off of this foundation, modifying either $\rho_\infty$ or $N$ to probe different portions of the spectrum with the same accuracy.  
%

\noindent The authors thank D. Griffiths and M. Kiessling for valuable commentary on an early draft.

\bibliographystyle{elsarticle-num}
\bibliography{BIH}
%
%
%
%
%
%
%

\end{document}